# Interfacial superconductivity and a Se-vacancy ordered insulating phase in the FeSe/PbO$_x$ heterostructures

Yunkai Guo[1,2], Xuanyu Long[3], Jingming Yan[1,2], Zheng Liu[4], Qi-Kun Xue[1,2,5,6*] and Wei Li[1,2,7†]

[1]*State Key Laboratory of Low-Dimensional Quantum Physics, Department of Physics, Tsinghua University, Beijing 100084, China*

[2]*Frontier Science Center for Quantum Information, Beijing 100084, China*

[3]*Institute for Advanced Study, Tsinghua University, Beijing 100084, China*

[4]*School of Physics, Beihang University, Beijing 100191, China*

[5]*Beijing Academy of Quantum Information Sciences, Beijing 100193, China*

[6]*Southern University of Science and Technology, Shenzhen 518055, China*

[7]*Hefei National Laboratory, Hefei 230088, China*

**ABSTRACT**. The discovery of high-temperature superconductivity in FeSe/SrTiO$_3$ has sparked significant interests in exploring new superconducting systems with engineered interfaces. Here, using molecular beam epitaxy growth, we successfully fabricate FeSe/PbO$_x$ heterostructures and discover superconductivities in three different monolayer FeSe-related interfaces. We observe superconducting gaps of 13~14 meV in the monolayer FeSe films grown on two different phases of PbO$_x$. Moreover, we discover a new insulating Fe$_{10}$Se$_9$ phase with an ordered √5 × √5 Se-vacancy structure. Our first-principles calculation suggests that this new insulating phase originates from electronic correlation. Intriguingly, an additional monolayer FeSe film grown on the insulating Fe$_{10}$Se$_9$ also exhibits superconductivity with the gap size of 5 meV. Our results suggest that the work function differences between the monolayer FeSe and the substrates, which can induce band bending and charge transfer, are crucial for the interfacial superconductivity.

## I. INTRODUCTION.

The discovery of the interfacial superconductivity in monolayer FeSe film grown on SrTiO$_3$ (FeSe/STO) [1] has opened a new avenue for the development of high-temperature superconductors. The superconducting gap closing temperature ($T_{gc}$) in FeSe/STO can reach as high as 83 K [1-6]. In addition to superconductivity, various novel phases, such as nematicity [3,7] and stripe order [8,9], have also been observed in FeSe/STO.

The influence of different substrates on superconductivity has been extensively studied [10-15]. For instance, the $T_{gc}$ of monolayer FeSe on BaTiO$_3$ can reach up to 75 K [10,11]. In monolayer FeSe grown on anatase TiO$_2$, the measured superconducting gap can reach 21 meV, even slightly larger than that in FeSe/SrTiO$_3$ [12]. Enhanced superconductivity has also been reported in monolayer FeSe grown on the substrates other than TiO$_2$ terminations, such as MgO [13], GaO [14] and FeO [15]. Substrates play two key roles for the superconductivity. First, they provide sufficient charge carriers to monolayer FeSe and renormalize the band structures topology of FeSe [2,3,16]. Second, they offer an additional electron-phonon coupling channel, characterized by the shake-off bands observed in ARPES measurements [10,17]. However, the mechanism of charge transfer from the substrate to FeSe remains controversial. Possible mechanisms include oxygen vacancies-induced doping [18] and work function-related band-bending effects [19].

As a typical band insulator, α-PbO shares an identical crystal structure with FeSe, but its in-plane lattice constants ($a = b = 3.99$ Å) are larger than those of both STO and FeSe. Here, we use molecular beam epitaxy to grow monolayer PbO$_x$ on STO(001), and subsequently deposit few-layer FeSe films on PbO$_x$. Enhanced superconductivities are observed in monolayer FeSe/PbO$_x$. Additionally, we discover a new insulating Fe$_{10}$Se$_9$ phase with an ordered √5 × √5 Se-vacancy structure. Interestingly, superconducting gaps are also observed in an additional monolayer FeSe film grown on √5 × √5 Fe$_{10}$Se$_9$. These findings offer further insight into the charge transfer mechanism in monolayer FeSe-related systems.

## II. EXPERIMENTAL METHODS

*Contact author: qkxue@mail.tsinghua.edu.cn

†Contact author: weili83@tsinghua.edu.cn

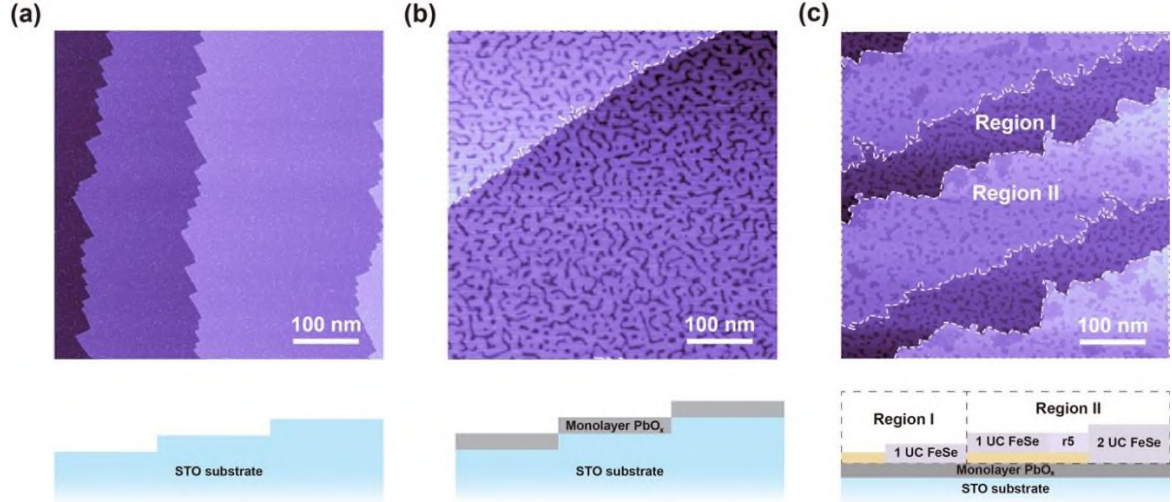

FIG. 1. The growth of FeSe/PbO$_x$ on SrTiO$_3$(001). (a) STM topographic image of treated SrTiO$_3$(001) substrate (500 nm × 500 nm, $V_b$ = 2 V, $I_t$ = 20 pA). Flat substrate terraces can be observed. The lower panel is the schematic side view of STO. The shape of the steps corresponds to the terraces of STO. (b) STM topography of an insulating PbO$_x$ monolayer film on SrTiO$_3$(001) (500 nm × 500 nm, $V_b$ = 7 V, $I_t$ = 20 pA). Its thickness is 400 pm. The step indicated by white dashed line in the topography is the terrace of the STO itself. The lower panel is the side view of PbO$_x$/STO. (c) Two distinct regions of monolayer FeSe/PbO$_x$ (500 nm × 500 nm, $V_b$ = 2 V, $I_t$ = 20 pA). The lower panel is the side view of FeSe/PbO$_x$. Two regions are separated by black dotted lines.

The Nb-doped (0.05 wt%) STO (001) substrates were degassed in an ultrahigh vacuum chamber (base pressure is better than $3 \times 10^{-10}$ Torr) at 500 °C for several hours and subsequently annealed at 1150 °C for 20 min to obtain TiO$_2$ terminated surface. To prepare PbO$_x$ films, STO substrates were kept at 450 °C, and high purity Pb (99.999%) sources were evaporated by a Knudsen cell. High purity Fe (99.995%) and Se (99.9999%) sources were co-evaporated by two Knudsen cells to grow FeSe films. During the growth, the PbO$_x$/STO were kept at 430 °C. The as-grown samples were annealed at 430 °C for one hour to improve the sample quality.

In-situ STM measurements were performed at 4.8 K in a commercial STM (Unisoku). A polycrystalline PtIr STM tip was calibrated on an Ag island before STM experiments. STS data were taken by a standard lock-in method. The feedback loop is disrupted during data acquisition with the frequency of oscillation signal of 973.0 Hz.

All of the DFT calculations are obtained by using the Vienna ab initio Simulation Package (VASP)[20]. A freestanding single layer of Fe$_{10}$Se$_9$ unit cell with $\sqrt{5} \times \sqrt{5}$ Se vacancy order is constructed with STM measured in-plane lattice constant 8.7 Å and a vacuum layer of 15 Å. All of the internal ion positions are relaxed until the residual forces are less than $10^{-2}$ eV/Å. The plane-wave cutoff is 500 eV in combination with the projector augmented wave (PAW) method [21]. The Monkhorst-Pack k-point grid [22] is $8 \times 8 \times 1$. The exchange and correlation is treated by using the Perdew-Burke-Ernzerhof generalized radient approximation (GGA) functional [20]. The convergence criteria is $10^{-5}$ eV for electronic iterations. The rotational invariant $+U$ correction introduced by Dudarev et al. [21] with $U$ = 2 eV is used for the DFT+U calculations. For the DOS calculation, the tetrahedron method with Blöchl corrections [22] is used with a $16 \times 16 \times 1$ k-point grid. Afterwards, the DOS is smoothened by convolution with an 8 meV-variance Gaussian function.

### III. RESULTS

Figure 1(a) shows the topography of treated STO, revealing large, flat terraces. To aid in understanding the sample preparation, schematic side views are provided in the lower panel of Fig. 1. Oxygen atoms emitted by heating STO [12,23] combine with Pb atoms evaporated from a Knudsen cell, forming monolayer PbO$_x$ on STO surface [see Fig. 1(b)]. By adjusting the temperatures of both the substrate and the Pb source, we can achieve the coverage of the monolayer PbO$_x$ film up to 72%. The details for the growth process are shown in Fig. S1[24]. The height of the monolayer PbO$_x$ is 400 pm. The band gap size of PbO$_x$ can be various, ranging from 1.5 to 4.2 eV [25-28]. For our STM measurements, stable scanning is only achievable when the bias voltage is higher than 7 V. Obtaining atomic resolution is challenging due to the insulating nature of the PbO$_x$ film. Subsequently, FeSe films are grown on top of the PbO$_x$ layer [Fig.1(c)]. Two distinct regions are observed in FeSe/PbO$_x$ [see Fig. 1(c)]. Region I is mostly monolayer FeSe, while Region II is more complex and will be discussed in detail later.

*Contact author: qkxue@mail.tsinghua.edu.cn

†Contact author: weili83@tsinghua.edu.cn

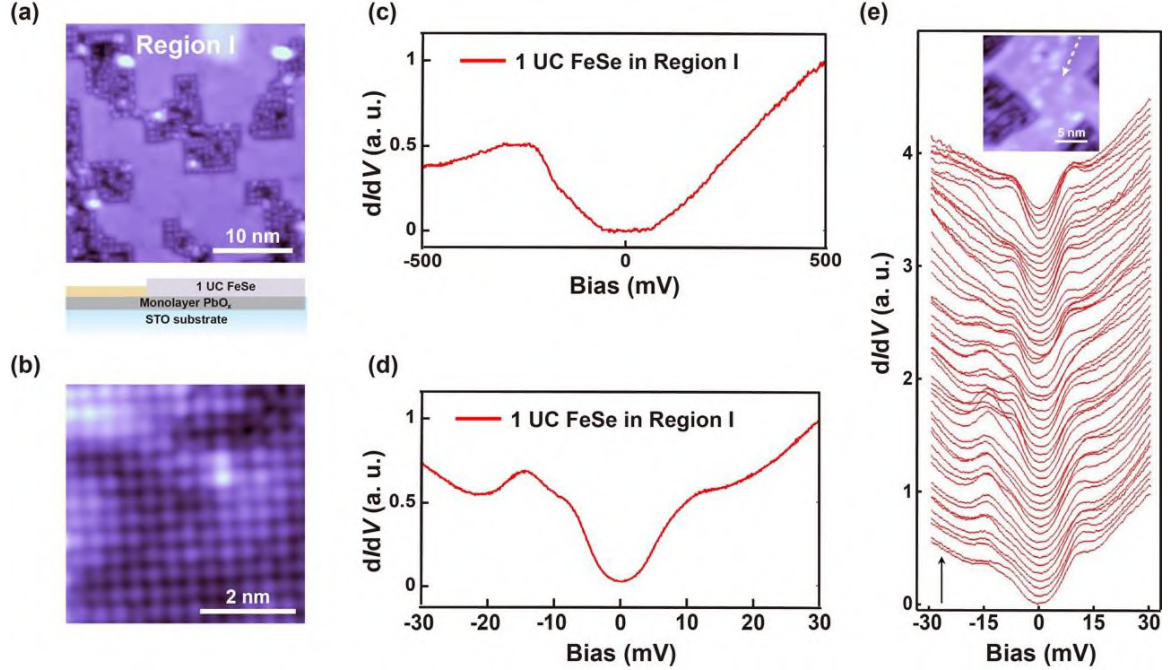

FIG. 2. Superconducting monolayer FeSe film grown on PbO$_x$. (a) STM topographic image of Region I (30 nm × 30 nm, $V_b$ = 2 V, $I_t$ = 20 pA). The darker and rougher areas are another insulating phase, while the flatter areas are the monolayer FeSe. The lower panel is the schematic side view of Region I. The dark and rough areas are in yellow, while 1UC FeSe are in purple. (b) Atomically resolved STM topography of monolayer FeSe in Region I (5 nm × 5 nm, $V_b$ = 100 mV, $I_t$ = 20 pA). (c) Large-scale d$I$/d$V$ spectrum of the monolayer FeSe in Region I ($V_b$ = 500 mV, $I_t$ = 200 pA). (d) Typical low energy d$I$/d$V$ spectrum ($V_b$ = 30 mV, $I_t$ = 200 pA). The superconducting gap is about 14 meV. (e) A series of low-energy d$I$/d$V$ spectra ($V_b$ = 30 mV, $I_t$ = 200 pA) taken along the white arrow in inset. Inset: STM topography of Region I (18 nm × 18 nm, $V_b$ = 2 V, $I_t$ = 20 pA).

We first focus on Region I. Its topography and schematic side view are shown in Fig. 2(a). The flat areas consist of monolayer FeSe grown on PbO$_x$, where tetragonal lattice is observed [Fig. 2(b)]. Since STO is not completely covered by PbO$_x$, some rougher and insulating areas [the lower parts in Fig. 2(a)] are visible and they may have different stoichiometry, as indicated by the yellow sections in the schematic in the lower panel of Fig. 2(a). The STS looks flat and lacks DOS for a large bias range ±60 mV, consistent with prior literature reports [11,29-33]. This indicates that the hole pocket near the Γ point lies below the Fermi level, suggesting that electrons transfer from the substrate to 1 UC FeSe [2]. The superconducting gap of monolayer FeSe/PbO$_x$ can reach 14 meV [Fig. 2(d)]. Due to the limited quality of the FeSe/PbO$_x$ interface, the gaps fluctuate [Fig. 2(e)].

Now we focus on Region II, where four different FeSe phases are observed [Fig. 3(a)]. The areas labeled as 2 unit cell (UC) FeSe are 550 pm (the height of 1 UC FeSe) higher than the monolayer FeSe in Region I, indicating that they are 2 UC FeSe grown on the PbO$_x$ [see the schematic in the lower panel of Fig. 3(a)]. The electronic structure of 2 UC FeSe is similar to that of 2 UC FeSe/STO (Fig. S2 [24]). The stripe charge order is rather weak due to the small size of the nematic domains (Fig. S3 [24]). In the mixture areas, 1 UC and 2 UC FeSe are interspersed (see Fig. S4 [24]).

The 1 UC FeSe in Region II is positioned higher than the 1 UC FeSe in Region I. As a result, the 1 UC FeSe here (purple layer) is located on the insulating layer mentioned earlier [the yellow layer in the schematic side views in Fig. 1(c) and 3(a)]. Notably, the 1 UC FeSe here also exhibits superconductivity, with a gap size of 13 meV [Fig. 3(g) and Fig. 3(h)].

*Contact author: qkxue@mail.tsinghua.edu.cn

†Contact author: weili83@tsinghua.edu.cn

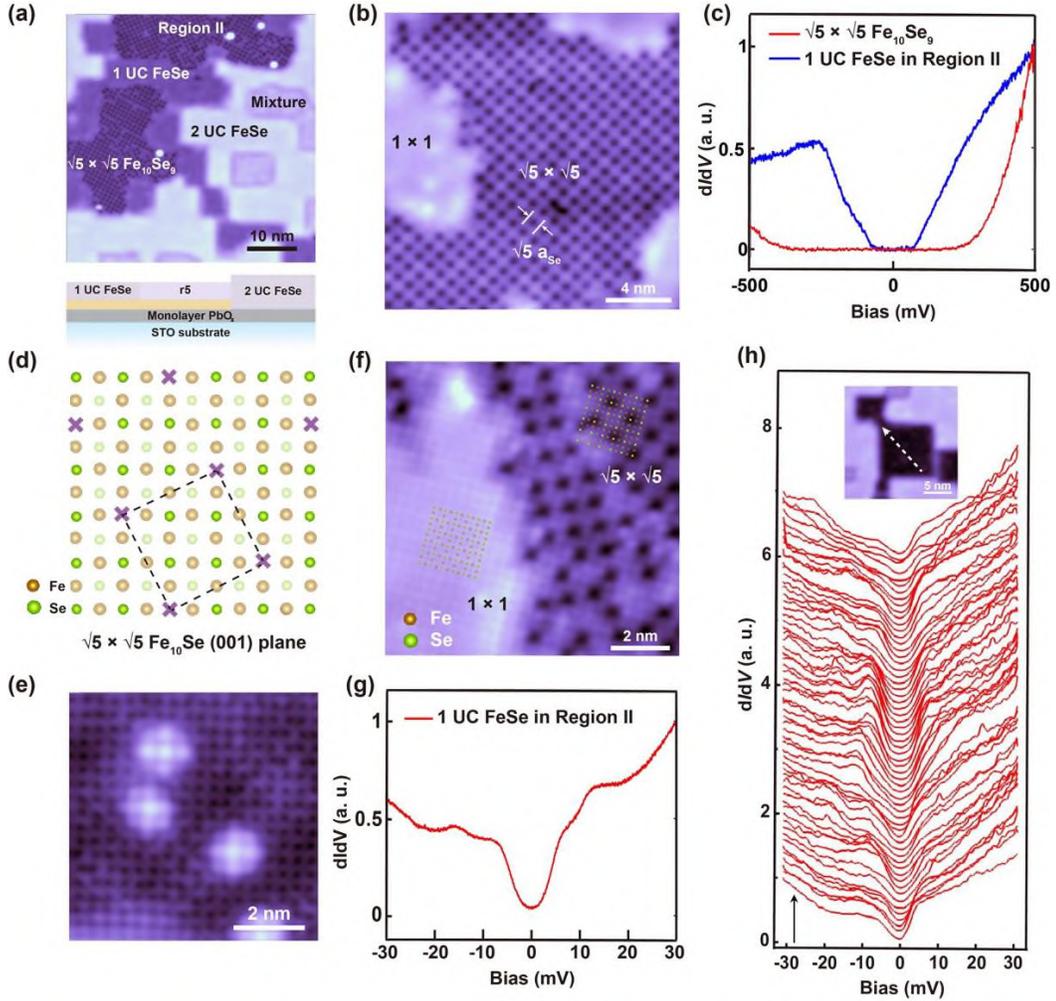

FIG. 3. Different phases in Region II. (a) STM topography of Region II (50 nm × 50 nm, $V_b$ = 2 V, $I_t$ = 20 pA), in which four types of FeSe can be observed. The lower panel is the schematic side view of Region II. The "r5" represents $\sqrt{5} \times \sqrt{5}$ Fe$_{10}$Se$_9$. The mixture phase is omitted in the schematic. (b) Zoom-in STM topography of $\sqrt{5} \times \sqrt{5}$ Fe$_{10}$Se$_9$ and 1 × 1 FeSe in Region II (18 nm × 18 nm, $V_b$ = 3.5 V, $I_t$ = 20 pA). The grid-like morphology corresponds to the $\sqrt{5} \times \sqrt{5}$ Fe$_{10}$Se$_9$ structure, while the flat surface corresponds to FeSe. The spacing between the grids is $\sqrt{5}$ times the lattice constant of Se $a_{Se}$. (c) d$I$/d$V$ spectrum ($V_b$ = 500 mV, $I_t$ = 200 pA) of the $\sqrt{5} \times \sqrt{5}$ and 1 × 1 phase. Compared to the 1 × 1 phase, the $\sqrt{5} \times \sqrt{5}$ phase exhibits significantly pronounced insulating characteristics. (d) The structure of the $\sqrt{5} \times \sqrt{5}$ Se-vacancy pattern as seen from (001) planes. The topmost Se atoms are in green, while the lowermost Se atoms are in light green. The middle Fe atoms are in brown. The positions of Se vacancies are marked by purple crosses, which form $\sqrt{5} \times \sqrt{5}$ order. A unit cell is marked by the black dashed lines, which means that $\sqrt{5} \times \sqrt{5}$ Fe$_{10}$Se$_9$ phase corresponds to Fe$_{10}$Se$_9$. (e) Atomically resolved STM topography of monolayer FeSe in region II (7 nm × 7 nm, $V_b$ = -10 mV, $I_t$ = 20 pA). Defects can be observed. (f) Atomically resolved STM topography of monolayer FeSe and $\sqrt{5} \times \sqrt{5}$ Fe$_{10}$Se$_9$ (10 nm × 10 nm, $V_b$ = -3 V, $I_t$ = 20 pA) on Region II. 1 × 1 monolayer FeSe phase and $\sqrt{5} \times \sqrt{5}$ Fe$_{10}$Se$_9$ phase coexist. In the superposed schematic lattice, Fe atoms are in brown, while Se atoms are in green. Square lattices are formed by the topmost Fe and Se atoms. Se-vacancies (the black holes) form the $\sqrt{5} \times \sqrt{5}$ order. (g) Typical low energy d$I$/d$V$ spectrum ($V_b$ = 30 mV, $I_t$ = 200 pA). The superconducting gap is about 13 meV. (h) A series of spectra taken along the white arrow in inset ($V_b$ = 30 mV, $I_t$ = 200 pA). The superconductivity is not uniform due to the sample quality. Inset: STM topography of a large area of monolayer FeSe in region II (20 nm × 20 nm, $V_b$ = 2.5 V, $I_t$ = 20 pA). The dark areas are monolayer FeSe, while the light areas are 2 UC FeSe.

*Contact author: qkxue@mail.tsinghua.edu.cn

†Contact author: weili83@tsinghua.edu.cn

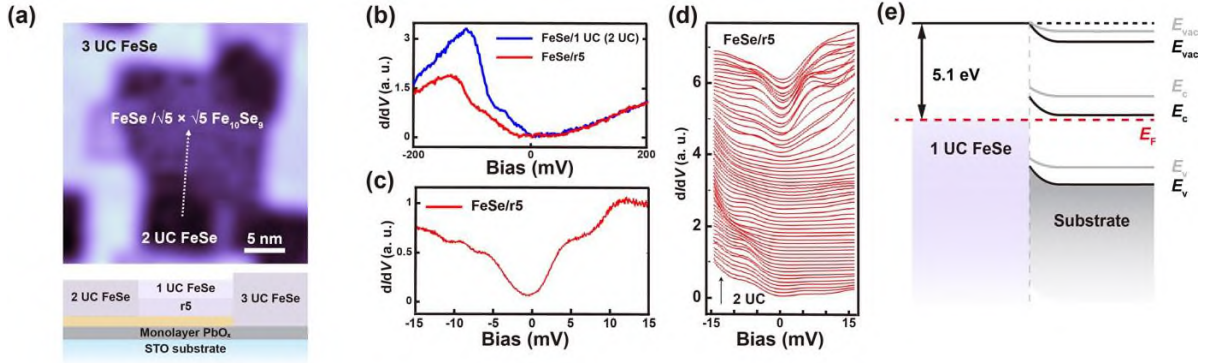

FIG. 4. Topography and electronic structures of additional monolayer FeSe layers grown on Region II. (a) STM topography (30 nm × 30 nm, $V_b$ = 2 V, $I_t$ = 20 pA) after depositing additional FeSe layers on Region II, in which three types of FeSe (2 UC FeSe, FeSe/√5 × √5 $Fe_{10}Se_9$ and 3 UC FeSe) can be observed. The schematic diagram in lower panel is the side view of FeSe/Region II. The "r5" represents √5 × √5 $Fe_{10}Se_9$. (b) Large-scale d$I$/d$V$ spectra of the 2 UC and FeSe/√5 × √5 $Fe_{10}Se_9$ phase ($V_b$ = 500 mV, $I_t$ = 200 pA). The d$I$/d$V$ of FeSe/r5 has a flatter bottom near Fermi level than that in 2 UC FeSe. (c) Small-scale d$I$/d$V$ spectrum ($V_b$ = 15 mV, $I_t$ = 200 pA) of the FeSe/√5 × √5 $Fe_{10}Se_9$ phase, showing superconducting-like behaviors. (d) A series of d$I$/d$V$ spectra with low energy ranges taken along the white arrow in **a** ($V_b$ = 15 mV, $I_t$ = 200 pA). A gap feature can be observed in FeSe/√5 × √5 $Fe_{10}Se_9$ phase. (e) The band-bending effect in the interfaces between 1 UC FeSe and different substrates. The bands denoted by black lines have more remarkable band-bending effect than the gray ones.

In addition, we identified a new insulating phase in Region II, denoted as $Fe_{10}Se_9$, with an ordered √5 × √5 Se-vacancy structure. In Figure 3(b), the grid-like pattern is associated with the √5 × √5 $Fe_{10}Se_9$ structure, whereas the flat surface is characteristic of FeSe. Atomic-resolution images of both 1 UC FeSe and √5 × √5 $Fe_{10}Se_9$ are obtained in Fig. 3(f). The left side of Fig. 3(f) displays the 1 × 1 Se-lattice of 1 UC FeSe. While on the right side, the black holes correspond to the Se-vacancies arranged in a √5 × √5 order. In the schematic plot [Fig. 3(d)], purple crosses indicate the Se-vacancies, with the distance between the nearest vacancies √5 times the lattice constant of Se atoms. In contrast to the 1 × 1 phase, a large insulating gap is observed in the tunneling spectrum of the √5 × √5 phase [Fig. 3(c)]. As shown in Fig. S5 of the Supplemental Material [24], the $Fe_{10}Se_9$ phase exhibits distinct vacancy characteristics, contrasting with prior reports [34]: (1) Se vacancies form stable √5 × √5 ordered structures under varying bias voltages [Fig. S5(a)], confirming their morphological origin rather than electronic effects; (2) Isolated Fe vacancies are also observed (Fig. S5(b) and (c), where missing Fe atoms (red markers) expose their bonded topmost Se atoms. These dual signatures unambiguously identify the $Fe_{10}Se_9$ phase.

We deposit an additional monolayer FeSe on Region II, and the original 1 UC FeSe and √5 × √5 $Fe_{10}Se_9$ become 2 UC FeSe and 1 UC FeSe/√5 × √5 $Fe_{10}Se_9$, respectively [Fig. 4(a)]. Large-scale d$I$/d$V$ spectra of the 2 UC FeSe and 1 UC FeSe/√5 × √5 $Fe_{10}Se_9$ are of similar shape [Fig. 4(b)], while the DOS of FeSe/√5 × √5 $Fe_{10}Se_9$ near $E_F$ is flatter, which has similar tendency as the spectra shown in Fig. 2(c) and Fig. 3(c), indicating higher electron doping. Consistently, low-energy spectrum in 1UC FeSe/√5 × √5 $Fe_{10}Se_9$ shows a superconducting-like gap [Fig. 4(c)]. The tunneling spectra [Fig. 4(d)] taken along the white arrowed dashed line in Fig. 4(a) indicate that the gap-like feature can be only observed in FeSe/√5 × √5 $Fe_{10}Se_9$, but absent in 2 UC FeSe (i.e., 1 UC FeSe/FeSe).

## IV. DISCUSSION AND CONCLUSIONS

We now discuss the potential origin of the charge transfer to monolayer FeSe. In total, we have discovered superconductivities in three monolayer FeSe-related interfaces. The observed DOS suppression is attributed to a superconducting gap as supported by the following evidence detailed in the Supplementary Materials [24]: (1) magnetic field-induced gap suppression in Regions I/II [Fig. S6 (d)–(e)]; (2) zero-bias current nulling (−1.5 to +1.5 mV) in $I$-$V$ curves [Fig. S6 (a)-(c)]; (3) persistent DOS reduction contrasting with non-superconducting 2 UC FeSe/ rough $PbO_x$ [Fig. S6(c), (f)]; (4) STS signatures mirroring superconducting 1 UC FeSe/STO, including ±60meV DOS depletion [Figs. 2(c), 3(c) and 4(b)]. Since the detailed lattice structures of the three interfaces are distinct, we suggest that the work function differences and the consequent band bending between the monolayer FeSe and the adjacent substrates play leading roles for the charge transfer. The work function of FeSe is 5.1 eV [19], while the work function of $PbO_x$ varies from 4.8 to 5.5 eV, depending on the growth conditions [35]. In our study, Pb atoms are evaporated using a Knudsen cell, while oxygen atoms are supplied by the heating of STO. This process results in an excess of Pb, thereby giving rise to the formation of n-type $PbO_x$. Consequently, the work function of n-type $PbO_x$ is lower than that of FeSe, allowing electrons to transfer from $PbO_x$ to FeSe [Fig. 4(e)]. This is the case for the monolayer FeSe in Region I. For the monolayer FeSe on the rougher $PbO_x$ area in Region II, comparable charge transfer [see Fig. 2(c) and Fig. 3(c)] and superconducting gaps [see Fig. 2(d) and Fig. 3(g)] are observed, indicating that the two types of $PbO_x$ have similar work functions and the band banding is denoted by the black lines in Fig. 4(e). For the additional monolayer FeSe layers grown on $\sqrt{5} \times \sqrt{5}$ $Fe_{10}Se_9$, the observed gap size is smaller, suggesting a smaller amount of charge transfer to FeSe, and a higher work function of the insulating $\sqrt{5} \times \sqrt{5}$ $Fe_{10}Se_9$. And the corresponding band bending is denoted by the gray lines in Fig. 4(e). The hypothesis of charge transfer can be further corroborated by large-scale $dI/dV$ spectra. Spectral hump positions (−240 meV in Region I, −220 meV in II, −140 meV in FeSe/$\sqrt{5}\times\sqrt{5}$, −130 meV in 2 UC FeSe; Fig. S7 [24]) track electron doping levels via Fe $3d_{z^2}$-band shifts near the Γ point [3]. Stronger doping in Regions I/II (humps farthest from $E_F$) correlates with larger superconducting gaps, while slightly doping in FeSe/$\sqrt{5}\times\sqrt{5}$ yields a reduced "gap-like" feature.

*Contact author: qkxue@mail.tsinghua.edu.cn

†Contact author: weili83@tsinghua.edu.cn

An insulating Se-deficient FeSe structure is not known before. A previous first-principles study on FeSe monolayer with artificial 2×2 Se-vacancy ordering shows a metallic band structure within the framework of standard density functional theory (DFT) [36]. Our own DFT calculation (shown in the Supplemental Material [24]) based on the experimental $\sqrt{5} \times \sqrt{5}$ $Fe_{10}Se_9$ superstructure does not find an insulating ground state either [see Figs. S8(a)-(b)]. Nevertheless, introducing antiferromagnetic (AFM) orders to the Fe plane is found to drastically suppress the density of states at the Fermi level, and an additional +U correction to the Fe 3d-orbitals clearly opens a gap [see Figs. S8(c)-(f)]. We have tested both the Néel order [see Fig. S8(c)] and a less regular AFM configuration by arbitrarily assigning 5 Fe spins within the supercell pointing up and the rest pointing down [See Fig. S8(e)]. The gap opening behaviors are qualitatively the same. However, without the AFM order, the +U correction alone is insufficient to drive a metal-to-insulator transition [see Fig. S8(b)]. The scenario appears similar to FeSe with $\sqrt{5} \times \sqrt{5}$ ordered Fe vacancies [37], which has attracted extensive studies because tuning FeSe from a superconductor to an insulator provides a unique angle to understand iron-based superconductors [34,37-41]. Intuitively, the intact Fe plane in the new $\sqrt{5} \times \sqrt{5}$ $Fe_{10}Se_9$ structure bridges to the superconducting pristine FeSe in a more natural way. Our preliminary calculations suggest that electron correlation plays an important role in opening the insulating gap. The same correlation might also underpin the superconducting phase. While a thorough understanding calls for further study, we believe that this new insulating phase contains novel clues for a deeper insight and further optimization of high-temperature superconductivity.

ACKNOWLEDGMENTS

The research was supported by the National Science Foundation of China (Grants Nos. 92365201, 52388201, 12374062), the Ministry of Science and Technology of China (Grant No. 2022YFA1403100), and the Innovation Program for Quantum Science and Technology (2021ZD0302400).

*Contact author: qkxue@mail.tsinghua.edu.cn

†Contact author: weili83@tsinghua.edu.cn

*Contact author: qkxue@mail.tsinghua.edu.cn

†Contact author: weili83@tsinghua.edu.cn

*Contact author: qkxue@mail.tsinghua.edu.cn

†Contact author: weili83@tsinghua.edu.cn